\documentstyle[12pt]{article}
\begin{document}

\title{ Angular Momentum Surface Density of the Kerr Metric}
\author{L. Fern\'andez-Jambrina\\and\\F.J. Chinea\vspace{1cm}\\Departamento de
F\'{\i}sica Te\'orica II,\\Facultad de Ciencias F\'{\i}sicas,\\Universidad
Complutense \\28040-Madrid, Spain} \date{} \maketitle
\abstract{A method for interpreting discontinuities of the twist potential of
vacuum stationary axisymmetric solutions of Einstein's equations is introduced.
Surface densities for the angular momentum of the source can
be constructed after solving a linear partial differential equation with
boundary conditions at infinity. This formalism is applied to the Kerr metric,
obtaining a regularized version of the density calculated with other formalisms.
The main result is that the integral defining the total angular momentum is
finite for the Kerr metric.  }
\vspace{0.5cm}

\hfil
\newpage

The physical interpretation of solutions of the Einstein equations is a task of
great importance in General Relativity. In the last few decades a great effort
has been done in the generation of new solutions, but only a few have been
studied from the physical point of view. In this letter we provide a new method
for constructing the angular momentum density of a source of a vacuum
stationary axisymmetric solution of Einstein equations from the discontinuities
of its twist potential. This is a generalization to curved spacetime of
the potential theory formula for constructing dipole densities for a Newtonian
field. As an example, the developed formalism is applied
to the Kerr metric and the results are compared with those obtained by Israel
in \cite{is1}. A brief discussion is provided. 

  We
shall follow the formalism developed in \cite{ch}\cite{tes} for stationary
axisymmetric perfect fluids and consider vacuum as a rigidly rotating perfect
fluid whose four-velocity one-form $u$ is just the timelike leg of an
orthonormal spacetime vierbein. In canonical coordinates $(t,\phi,\rho,z)$, the
metric is written:
\begin{eqnarray}ds^{2}=-e^{2U}(dt+Ad\phi)^{2}+e^{-2U}[e^{2k}(d\rho^{2}+dz^{2})+\rho^2
d\phi^{2}]\label{eq:can}\end{eqnarray}
 and so we choose $u=e^{U}(dt+Ad\phi)$ and define a
 form $w$ ($w=*\omega$, where $\omega$ is the vorticity form associated to
$u$ and $*$ denotes the two-dimensional Hodge dual in the space
orthogonal to the orbits of the Killing vectors
$\{\partial_t,\partial_\phi\}$) from \begin{eqnarray} du=a\wedge
u+w\wedge\theta^1\end{eqnarray} where $\theta^1$ is an one-form orthonormal to
$u$ in the space spanned by the orbits of the group of isometries, and $a$ is
the acceleration form. We shall only consider metrics which are asymptotically
flat. For our purposes we just need the following behavior at infinity in some
coordinates $(t,r,\theta,\phi)$:
\begin{eqnarray}ds^2=-(1-\frac{2m}{r})(dt+\frac{2J\sin^{2}\theta}{r}d\phi)^2+
\nonumber\\+(1+\frac{2m}{r})[dr^2+r^2(d\theta^2 +\sin^2\theta d\phi^2)]+O(1/r^2)
\end{eqnarray}
where $m$ is the total mass of the source and $J$ is the total angular momentum.
From the Bianchi and Einstein vacuum equations, it follows \cite{ch}\cite{tes}
that $w$ and $*w$ have to fulfil: 
\begin{eqnarray}dw+(b-2a)\wedge w=0 
\end{eqnarray} \begin{eqnarray}
d*w+2a\wedge *w=0
\end{eqnarray}
where $b=d(\ln\rho)$. 
These two equations can be formally integrated:
\begin{eqnarray}w=\rho^{-1}e^{2U}dA 
\end{eqnarray}
\begin{eqnarray}
*w=e^{-2U}d\chi
\end{eqnarray} 
 where $\chi$ is the so-called twist potential and $A$, the metric function in
(\ref{eq:can}). Einstein's equations for axisymmetric stationary vacuum metrics
can be reduced to a complex second order partial differential equation
for the Ernst potential \cite{Ernst}, $\varepsilon=e^{2U}+i\chi$. Our requirement
about the asymptotic form of the metric imposes the following condition on
$\chi$:
 \begin{eqnarray}
\chi=-2J\cos\theta/r^2+O(1/r^3)\end {eqnarray}

We shall integrate over the whole space $V_3$ orthogonal to the velocity $u$
(whose metric is $^3g=g+u\otimes u$) the projection of the gradient of a
function $z$ (to be determined later) over the difference $e^{-U}[*(w)-*w]$,
which will be obviously zero.  

\begin{eqnarray}
\sqrt{^3g}e^{-U}<[*(w)-*w],dz>=\sqrt{^3g}\partial_{\nu}z\{\rho^{-1}e^U\varepsilon^{\mu\nu}
\partial_{\mu}A-e^{-3U}g^{\mu\nu}\partial_{\mu}\chi\}\nonumber\\=\partial_{\mu}([\mu\nu]A
\partial_{\nu}z-\sqrt{^3g}e^{-3U}g^{\mu\nu}\chi\partial_{\nu}z)+\chi\partial_{\mu}
(\sqrt{^3g}e^{-3U}g^{\mu\nu}\partial_{\nu}z)
\end{eqnarray} 
where $\varepsilon^{\mu\nu}=e^{2(U-k)}[\mu\nu]$ is the Levi-Civit\`a tensor on
the space orthogonal to the orbits of the Killing vectors and $<\ ,\ >$ is the
scalar product associated to the metric. We choose $z$ so that it satisfies 

\begin{eqnarray}
\partial_{\mu}(\sqrt{^3g}e^{-3U}g^{\mu\nu}\partial_{\nu}z)=0\label{eq:lap}
\end{eqnarray} 

with the boundary condition $z=r\cos\theta$ at infinity. {\it We
assume now that the metric is continuous and that the twist potential is
discontinuous across a closed surface $S$.} This is no restriction since we can
always close the surface taking the zero value for the discontinuity in the rest
of it. The $3$-space $V_3$ is thus divided in two regions $V_3^-$ (interior)
and $V_3^+$ (exterior) and we can apply Stokes' theorem to our integral:
\begin{eqnarray}
0&=&\int_{V_3}\sqrt{^3g}e^{-U}<[*(w)-*w],dz>dx^1dx^2dx^3
\nonumber\\&=&\int_{\partial V_{3}^+\cup\partial
V_{3}^-}n_\mu\partial_{\nu}z\{\rho^{-1} e^U\varepsilon^{\mu\nu}A
-e^{-3U}g^{\mu\nu}\chi\}dS \end{eqnarray} 
The boundary of $V_{3}^+$ consists of $S$ and the sphere at infinity and the
boundary of $V_{3}^-$ is just $S$. According to our assumptions about
asymptotic behaviour and continuity, the integral at infinity can be computed to
yield:\begin{eqnarray}
\int_SdS[\chi]e^{-3U}g^{\mu\nu}n_\mu\partial_{\nu}z=-8\pi
J\label{eq:den}\end{eqnarray}where we denote by $[\chi]$ the jump of the twist
potential across $S$ and by $n$ the outward unit normal to $S$. From this
formula we can interpret
$-\frac{1}{8\pi}[\chi]e^{-3U}g^{\mu\nu}n_\mu\partial_{\nu}z$  as the angular
momentum density of the two-dimensional source of the vacuum metric.

 Now we can apply this result to an example: The metric discovered by Kerr
\cite{Kerr} is of great interest in astrophysics as the exterior of a rotating
black hole. In 1970, Israel \cite {is1} used the theory of surface layers
\cite{is2} to obtain the energy-momentum tensor of a minimal source for the Kerr
metric that consisted of a disk. The mass and angular momentum densities
obtained with this method are not integrable and therefore  the singularity ring
encircling the disk had to be considered. This was done by L\'opez applying the
theory of distributions \cite{lop}. In what follows we shall calculate the
angular momentum of the source taking into account the discontinuities of the
twist potential.

The Ernst potential for the Kerr metric is $\varepsilon=1-2m/(r-ia\cos\theta) $
and so the twist potential is just $\chi=-2ma\cos\theta/(r^2+a^2\cos^2 \theta)
 $. This function takes the value $\chi(r=0)=-2m\epsilon/(a\cos\theta) $,
$\theta\in[0,\pi/2)$, $\epsilon=\pm 1$, on approaching the disk $r=0$ from
above or from below, respectively. This can be easily seen \cite{is1}
considering pseudocylindrical coordinates:
\begin{eqnarray}
P=(r^2+a^2)^{1/2}\sin\theta
\end{eqnarray}
\begin{eqnarray}
Z=r\cos\theta
\end{eqnarray} 
The surface $r=0$ is then is a disk of radius $P=a\sin\theta$ and
therefore the points with polar angle $\theta$ and $\pi-\theta$ match on the
disk. Hence, in what follows we shall restrict the range of $\theta$ to
$[0,\pi/2)$ on the disk $r=0$ to avoid double-counting these points. The whole
metric has the following expression:

\begin{eqnarray}ds^2&=&-
(1-\frac{2mr}{r^2+a^2\cos^2\theta})(dt+\frac{2mar\sin^2\theta}{r^2+a^2\cos^2\theta}
d\phi)^2+\nonumber\\&+&(1-\frac{2mr}{r^2+a^2\cos^2\theta})^{-1}\{(r^2-2mr+a^2)\sin^2\theta
d\phi^2
+\nonumber\\&+&(r^2-2mr+a^2\cos^2\theta)(
\frac{dr^2}{r^2-2mr+a^2}+d\theta^2)\} 
\end{eqnarray}
where $m$ is the total mass and $ma$, the total angular momentum.
It fulfils the required asymptotic conditions for the previously described
formalism to be applied. A solution for (\ref{eq:lap}) with the prescribed
boundary condition at infinity is
\begin{eqnarray}
z=(r-3m)\cos\theta+\frac{2a^2m(5\cos^3\theta-3\cos\theta)}{5(r^2+a^2\cos^2\theta)}
\end{eqnarray} 
The disk $r=0$ has $n=\frac{1}{\cos\theta}\partial_r$ as a unit normal vector and
the metric on this surface is:

\begin{eqnarray}
ds^2=a^2(\cos^2\theta d\theta^2+\sin^2\theta d\phi^2)\end{eqnarray}

Inserting these expressions in (\ref{eq:den}) we get the following expression for
the angular momentum of the source:
\begin{eqnarray}
J=\int_0^{2\pi}d\phi\int_0^{\pi/2}d\theta a^2\sin\theta \cos\theta
\;\sigma_J
\end{eqnarray}
\begin{eqnarray}
\sigma_J=\frac{m}{2\pi a\cos\theta}\label{eq:reg}
\end{eqnarray} 
The integral defining $J$ is perfectly regular and takes the value $ma$, which
is consistent with the asymptotic expression of the metric.

The Ernst potential for the Kerr metric is a solution of the flat spacetime
Laplace equation in oblate spheroidal coordinates \cite{is1}. Therefore its
imaginary part can be viewed as a Newtonian potential and we can study its
dipole density using potential theory \cite {Kel}. As the disk of discontinuity
is flat, the expressions obtained for $J$ and $\sigma_J$ in flat spacetime
coincide with those obtained previously for the Kerr metric. Moreover if we
consider the classical potential that has $\sigma_J$ as a dipole source, it
happens to be the same as the twist potential. That is, the twist potential for
the Kerr metric, when considered as a Newtonian potential is completely
generated by a dipole layer. No higher multipole layers need to be considered.

We can also compare the density $\sigma_J$ with the one obtained in \cite{is1}.
In that reference, the calculated angular momentum density is
$\sigma=-m\sin^2\theta/4\pi a\cos^3\theta$. This density is negative and
non-integrable and so it has to be compensated with an infinite angular
momentum on the singularity ring. On the contrary, the density derived in this
letter is positive and integrable and so it could be viewed as a regularized
version of \cite{is1}. A similar situation happens in Magnetostatics, where the
potential due to a ring of uniform current can be obtained also from a constant
magnetic dipole density on the disk surrounded by the ring.

\noindent
{\it The present work has been supported in part by DGICYT Project PB89-0142;
L.F.J. is supported by a FPI Predoctoral Scholarship from Ministerio de
Educaci\'{o}n y Ciencia (Spain). The authors wish to thank L.M.
Gonz\'alez-Romero for valuable discussions.}

 \end{document}